%% file: conference_101719.tex
\documentclass[conference]{IEEEtran}

\usepackage{multirow}
\usepackage[table,dvipsnames]{xcolor} 
\usepackage{cite}
\usepackage{amsmath,amssymb,amsfonts}
\usepackage{graphicx}
\usepackage{textcomp}
\usepackage{listings}
\usepackage{algorithm}
\usepackage{algpseudocode}
\usepackage{xcolor}
\usepackage{subcaption}
\usepackage{MnSymbol}

\usepackage{makecell}
\usepackage{threeparttable}
\usepackage{enumitem}
\usepackage{mathtools}

\definecolor{molex}{HTML}{F7965A}
\definecolor{zinergy}{HTML}{00B0F0}
\definecolor{bluespark}{HTML}{0071BC}
\definecolor{harvester}{HTML}{009B55}

\def\BibTeX{{\rm B\kern-.05em{\sc i\kern-.025em b}\kern-.08em
    T\kern-.1667em\lower.7ex\hbox{E}\kern-.125emX}}
    
\begin{document}
\bstctlcite{IEEEexample:BSTcontrol} 
\setlength{\abovedisplayskip}{2ex}
\setlength{\belowdisplayskip}{2ex}

\title{Hardware-Aware Automated Neural Minimization for Printed Multilayer Perceptrons\vspace{0.5cm}}

\author{\IEEEauthorblockN{
Argyris Kokkinis\IEEEauthorrefmark{1},
Georgios Zervakis\IEEEauthorrefmark{4},
Kostas Siozios\IEEEauthorrefmark{1},
Mehdi B. Tahoori\IEEEauthorrefmark{3},
and J{\"o}rg Henkel\IEEEauthorrefmark{3}
}
\IEEEauthorblockA{\IEEEauthorrefmark{1}Aristotle University of Thessaloniki, Greece,
\IEEEauthorrefmark{4}University of Patras, Greece,
\IEEEauthorrefmark{3}Karlsruhe Institute of Technology, Germany}
\IEEEauthorblockA{
\IEEEauthorrefmark{1}\{arkokkin, ksiop\}@auth.gr,
\IEEEauthorrefmark{4}zervakis@ceid.upatras.gr,
\IEEEauthorrefmark{3}\{mehdi.tahoori, henkel\}@kit.edu
}
}

\maketitle

\begin{abstract}
The demand of many application domains for flexibility, stretchability, and porosity cannot be typically met by the silicon VLSI technologies. Printed Electronics (PE) has been introduced as a candidate solution that can satisfy those requirements and enable the integration of smart devices on consumer goods at ultra low-cost enabling also in situ and on-demand fabrication. However, the large features sizes in PE constraint those efforts and prohibit the design of complex  ML circuits due to area and power limitations. 
Though, classification is mainly  the core task in printed applications.
In this work, we examine, for the first time, the impact of neural minimization techniques, in conjunction with bespoke circuit implementations, on the area-efficiency of printed  Multilayer Perceptron classifiers.
Results show that for up to 5\% accuracy loss up to 8x area reduction can be achieved.
\end{abstract}

\begin{IEEEkeywords}
Approximate Computing, Multilayer Perceptrons, Neural Minimization, Printed Electronics
\end{IEEEkeywords}

\input{Sections/section1.tex}
\input{Sections/section2.tex}
\input{Sections/section3.tex}

\input{Sections/section4.tex}
\input{Sections/acknowledgment}
\input{Sections/references}

\end{document}

%% file: Sections/section1.tex
\section{Introduction \& Motivation}

Printed Electronics (PE) can deliver flexible, lightweight and low-cost devices that can be integrated on every-day consumer goods enabling the smart services on low-end healthcare products, disposables, packaged foods, beverages etc.,~\cite{Mubarik:MICRO:2020:printedml}.
Despite their benefits, the large feature sizes of PE lead to the design of large and power demanding circuits that are infeasible to deploy on small surfaces and operate under tight battery requirements. Especially, Multi-Layer Perceptrons (MLPs) has been proven to be difficult to map on printed technologies without compromising either the model’s accuracy or the design’s feasibility~\cite{Mubarik:MICRO:2020:printedml, Armeniakos:DATE2022:axml}.
The design of fully customized, a.k.a bespoke, circuits in PE has been suggested as a possible solution to optimize area and power overheads~\cite{Bleier:ISCA:2020:printedmicro,Mubarik:MICRO:2020:printedml}. In bespoke implementations the model’s coefficients are hardwired in the circuit, enabling thus per model customization-optimization.  
This approach is realistically feasible only in PE due to their ultra-low manufacturing and non-recurrent engineering costs.

Neural minimization techniques are widely used in the Machine Learning (ML) domain and in Deep Neural Networks (DNNs) as a solution to compress networks and potentially increase their performance for some accuracy loss.
In this work, we examine, for the first time, the impact of neural minimization techniques on the design of printed MLP classifiers. 
Specifically, we discuss and evaluate the impact of quantization, pruning, and weight clustering on the area-efficiency of bespoke MLP implementations.
Our results show that the area of printed MLPs can decrease up to $\mathrm{8}$x for up to $\mathrm{5\%}$ accuracy loss.

%% file: Sections/section2.tex
\section{Neural Minimization Techniques}

The hardware impact of quantization, pruning and weight clustering is explored. The first two techniques aim to reduce bit-width and the number of the network’s weights while the third is used to minimize the number of the bespoke multipliers. 
In printed bespoke architecture a combination of them can potentially minimize the size of the MLPs and deliver area and power gains. 

\subsection{Quantization}
Quantization or precision scaling is one of the most widely used approximation techniques for neural networks~\cite{HenkelICCAD2022}, since it preserves the regularity of the compute patterns prevalent in ML models.
For moderate quantization levels the accuracy loss is negligible but for extremely low precision the accuracy drop might become significant. 
As shown in~\cite{Armeniakos:DATE2022:axml}, in bespoke implementations the ML circuit's area directly depends on the selected coefficients.
In bespoke MLPs, the values of the network’s weights and the bit representation of the inputs directly define the area-requirements of the required multipliers and consequently adders.
As a result, quantizing the network’s weights to low bit-width leads to more hardware-``friendly'' weights that result in smaller arithmetic units and thus, more area-efficient neurons.

\subsection{Pruning}
Structured and un-structured pruning are well-studied approaches to compress DNNs and potentially skip operations leading also to performance gains.
In structured pruning nodes or layers are removed from the model whereas in unstructured pruning the number of neural connections is decreased. 
The former is mainly preferred due to the direct performance gains albeit that the latter delivers mainly higher accuracy for similar sparsity.
On the other hand, bespoke circuits can seamlessly benefit from unstructured pruning since the weights are hardwired in the printed MLP implementation.
If a connection is removed, then the respective multiplier is directly removed from the circuit.
Moreover, the corresponding neuron needs fewer sum operands leading also to more area-efficient addition.

\subsection{Weight Clustering}
Weight clustering is also a compression technique that has been explored towards minimizing the memory requirements of DNNs~\cite{sharing:ICLR2016}.
In bespoke MLPs, weight clustering can be leveraged to reduce the area requirements.
By forcing weights of the same position (i.e., multiplied by the same input) to the same value, the product can be shared among many operations and the number of the required multiplier units decreases accordingly.
We employ the weight clustering of~\cite{sharing:ICLR2016} to enable multiplier-sharing at the hardware-level.

%% file: Sections/section3.tex
\section{Evaluation Analysis}\label{sec:eval}

\begin{figure}[t!]
\centering
    \centering
    \includegraphics[width=\columnwidth]{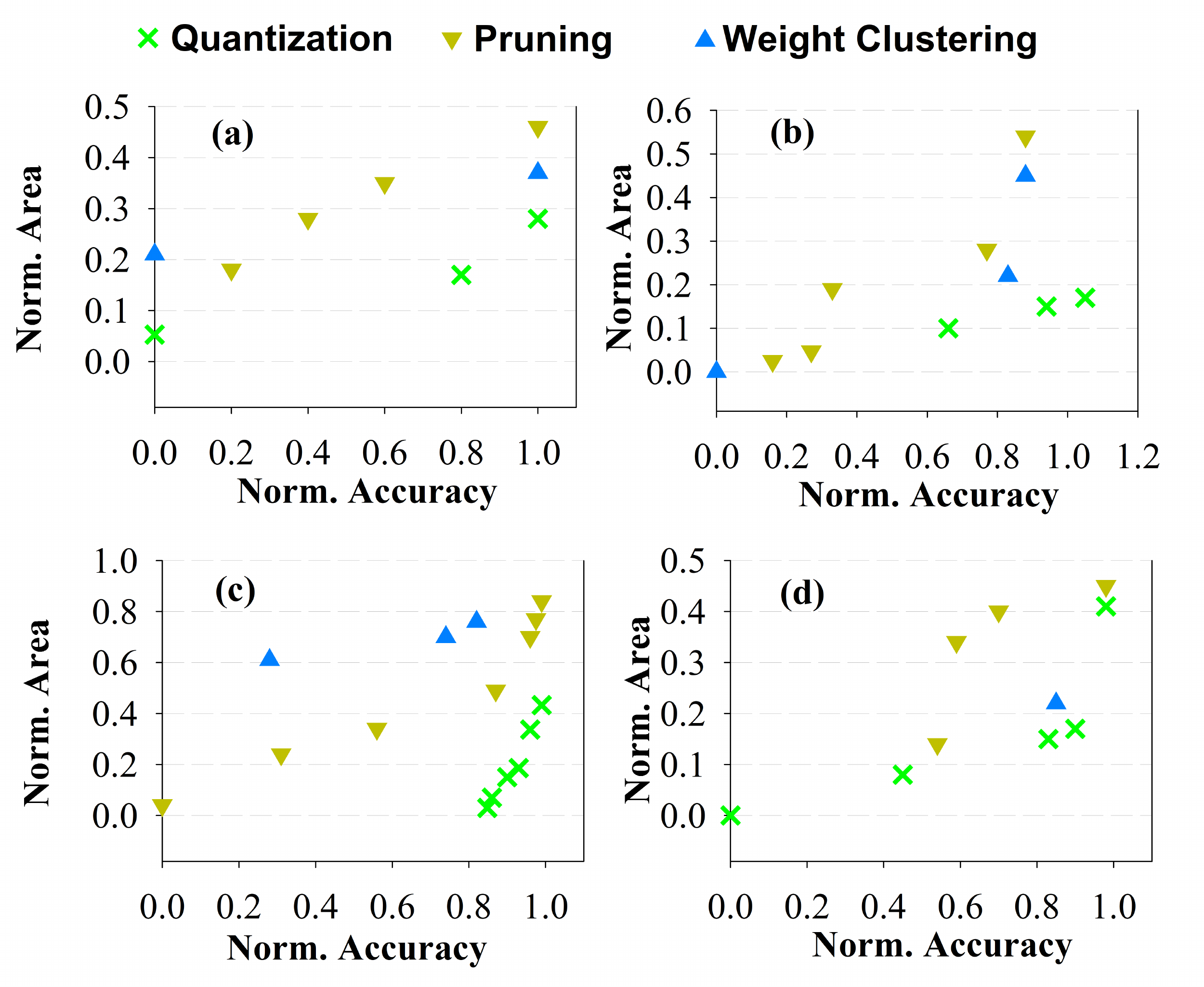} 
\caption{Area-Accuracy trade-off of the printed MLPs with quantization, pruning, and weight clustering~\cite{sharing:ICLR2016}. Values are normalized over each baseline MLP~\cite{Mubarik:MICRO:2020:printedml}.
Classifiers: (a) WhiteWine, (b) RedWine, (c) Pendigits, (d) Seeds.}
\label{fig:scatter_plots}
  \vspace{-.1in}
\end{figure}

\begin{figure}[t!]
\centering
    \centering
    \includegraphics[width=\columnwidth]{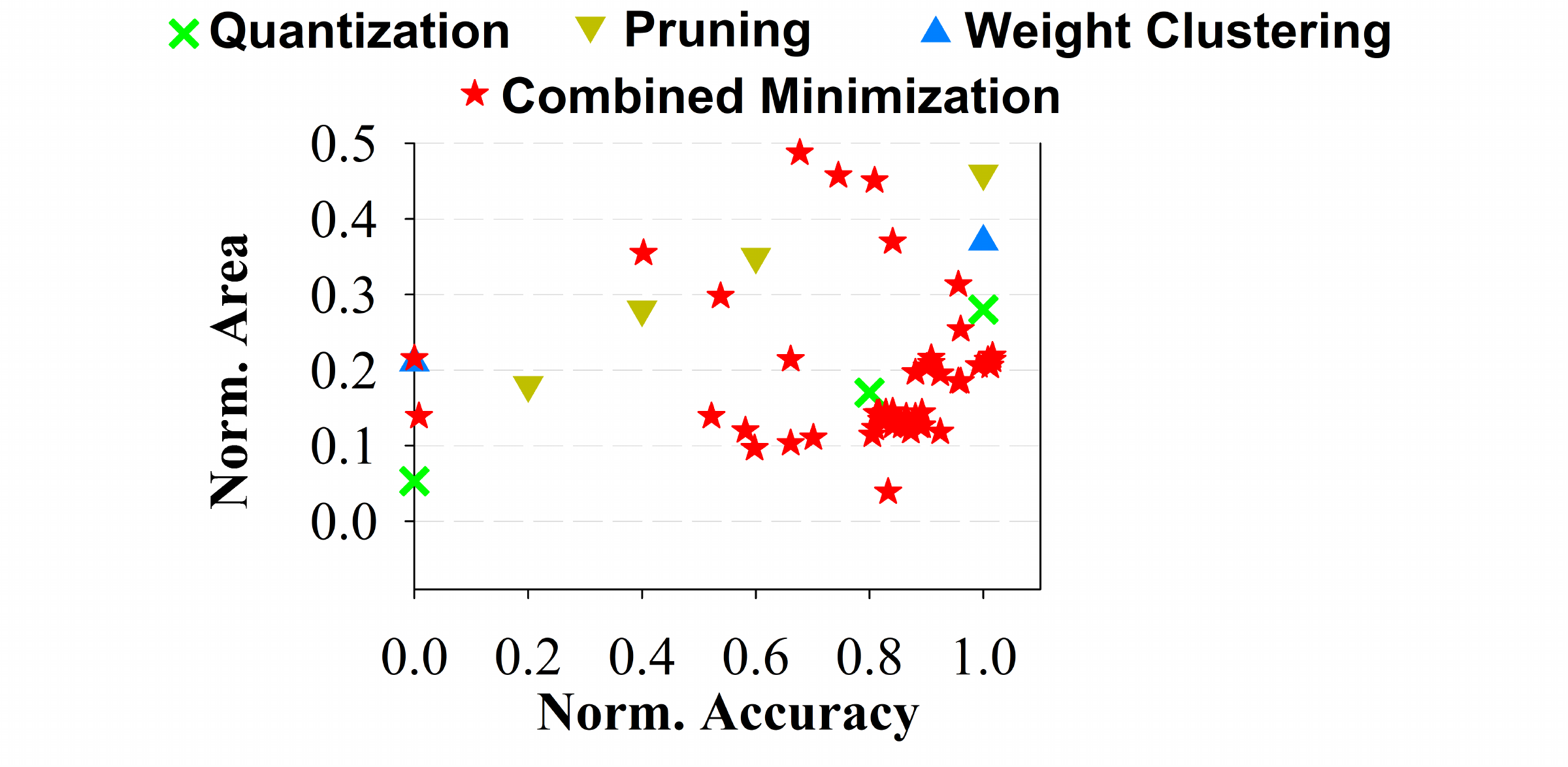} 
\caption{Area-Accuracy trade-off of the WhiteWine MLP classifier when quantization, pruning, weight clustering and all the three minimization techniques are combined.}
\label{fig:scatter_ww}
  \vspace{-.1in}
\end{figure}

In this section the efficiency of the aforementioned minimization techniques to reduce the area of printed MLPs is evaluated on four datasets: the WhiteWine, RedWine, Pendigits and Seeds datasets of the UCI ML repository~\cite{Dua:2019:uci}.
Qkeras~\cite{QKeras} is used to quantize the networks and a Quantization-Aware (re)-trainig (QAT) is performed. 
The Synopsys Design Compiler, the PrimeTime tools and the open source Electrolyte Gated Transistor (EGT) library~\cite{Bleier:ISCA:2020:printedmicro} are used for synthesis and hardware-analysis. 
The baseline in this evaluation analysis is the respective un-minimized bespoke MLP~\cite{Mubarik:MICRO:2020:printedml}.

Figure~\ref{fig:scatter_plots} shows the accuracy-area Pareto fronts when the three techniques are applied standalone on the examined classifiers. 
The axes on the Figure~\ref{fig:scatter_plots} plots are normalized  w.r.t the area-accuracy values of the baseline~\cite{Mubarik:MICRO:2020:printedml}.
In this evaluation unstructured pruning with a sparsity level between 20\% to 60\% is examined.
The quantization Pareto curves were generated by evaluating multiple designs with the bit precision of the classifiers’ quantized weights ranging between 2 to 7 bits. 
Finally, the weight clustering Pareto points were produced by executing the algorithm~\cite{sharing:ICLR2016} for a selected range of clusters. 
As expected all four minimization techniques lead to the generation of smaller classifiers that trade-off accuracy loss to area reduction.
The quantization Pareto front is better than the pruning and weight clustering, featuring on average 5x area reduction for up to 5\% accuracy loss, while for the same accuracy constraint the area gains when pruning and weight clustering techniques are applied are 2.8x and 3.5x respectively. Note, that the weight clustering approach generates MLPs that fullfil the 5\% accuracy threshold only for the RedWine and WhiteWine datasets. 

Finally, Figure~\ref{fig:scatter_ww} shows the area-accuracy trade-off when we combine all the three minimization techniques.
To obtain these designs we used a hardware-aware Genetic Algorithm.
As shown, the combination of all techniques leads to designs that feature high accuracy and lower area, outperforming the standalone techniques.
Interestingly, for the 5\% accuracy threshold the area gains reach up to 8x.

%% file: Sections/section4.tex
\section{Conclusions}
Printed electronics is a promising solution to enable smart services in application domains that have witnessed limited penetration of computing.
However, the high hardware overheads prohibit the realization of complex printed ML systems.
In this work, we showed how neural minimization techniques, initially designed for memory savings, may pave the way towards printed ML classification using MLPs.

%% file: Sections/acknowledgment.tex
\section*{Acknowledgments}
This work is supported by the E.C. Funded Programme ``SERRANO'' under H2020 Grant 101017168,
by the European Union (ERC, PRICOM, 101052764),
and by DFG through the project ``ACCROSS: Approximate Computing aCROss the System Stack'' HE 2343/16-1.

%% file: Sections/references.tex
\bibliographystyle{IEEEtran}
\bibliography{references}